\documentstyle[twoside,fleqn,espcrc2,epsf]{article}

\newcommand{\AmS}{{\protect\the\textfont2
  A\kern-.1667em\lower.5ex\hbox{M}\kern-.125emS}}

\def\spose#1{\hbox to 0pt{#1\hss}}
\def\ltapprox{\mathrel{\spose{\lower 3pt\hbox{$\mathchar"218$}}
 \raise 2.0pt\hbox{$\mathchar"13C$}}}
\def\gtapprox{\mathrel{\spose{\lower 3pt\hbox{$\mathchar"218$}}
 \raise 2.0pt\hbox{$\mathchar"13E$}}}
\def\inapprox{\mathrel{\spose{\lower 3pt\hbox{$\mathchar"218$}}
 \raise 2.0pt\hbox{$\mathchar"232$}}}

\newcommand{\be}{\begin{equation}}
\newcommand{\ee}{\end{equation}}

\def\su3{$SU(3)$}

\def\ham{{\bf \rm H}}

\title{Topological Susceptibility and Zero Mode Size in Lattice QCD}

\author{
  Robert~G.~Edwards\thanks{Speaker at the conference.
      This research was supported by DOE contracts
      DE-FG05-85ER250000 and DE-FG05-96ER40979.
      Computations were performed on the QCDSP, CM-2 and the 
      workstation cluster at SCRI.
      },
  Urs. M. Heller,
  and 
  Rajamani Narayanan
  \address{SCRI, Florida State University, 
      Tallahassee, FL 32306-4130, USA}
}
\begin{document}

\begin{abstract}

We use the overlap formalism to define a topological index on the
lattice. We study the spectral flow of the hermitian Wilson-Dirac
operator and identify zero crossings with topological objects.
We determine the topological susceptibility and zero mode size
distribution, and we comment on the stability of our results.
\end{abstract}

\maketitle

\section{INTRODUCTION}

In this proceedings, we summarize our work on the determination of the
topological susceptibility and zero mode size distribution of $SU(2)$
and $SU(3)$ gauge fields using the overlap formalism. Most of this
work has appeared in
print~\cite{su3_top}. We will focus here
on new results which are extensions of the published work, namely the
topological susceptibility for $SU(2)$ pure gauge theory, the zero mode size
distribution for both $SU(2)$ and $SU(3)$, and the density of zero
eigenvalues $\rho(0)$ for $SU(2)$ and $SU(3)$. 
%Comparisons are made
%with determinations 
%of these quantities 
%by other groups.

A more detailed discussion of the role of the zero eigenvalue density
can be found in the proceedings of the plenary talk given by
Narayanan~\cite{rnaray_lat98}. 
%The proceedings contribution of
%Heller~\cite{heller_lat98} describes ????.

\begin{figure*}[t]
\vspace*{-10mm} \hspace*{-0cm}
\begin{center}
\epsfxsize = \textwidth
%\centerline{\ewxy{b6.0_all.ps}{150mm}}
\centerline{\setlength{\epsfxsize}{150mm}\epsfbox[50 60 680 570]{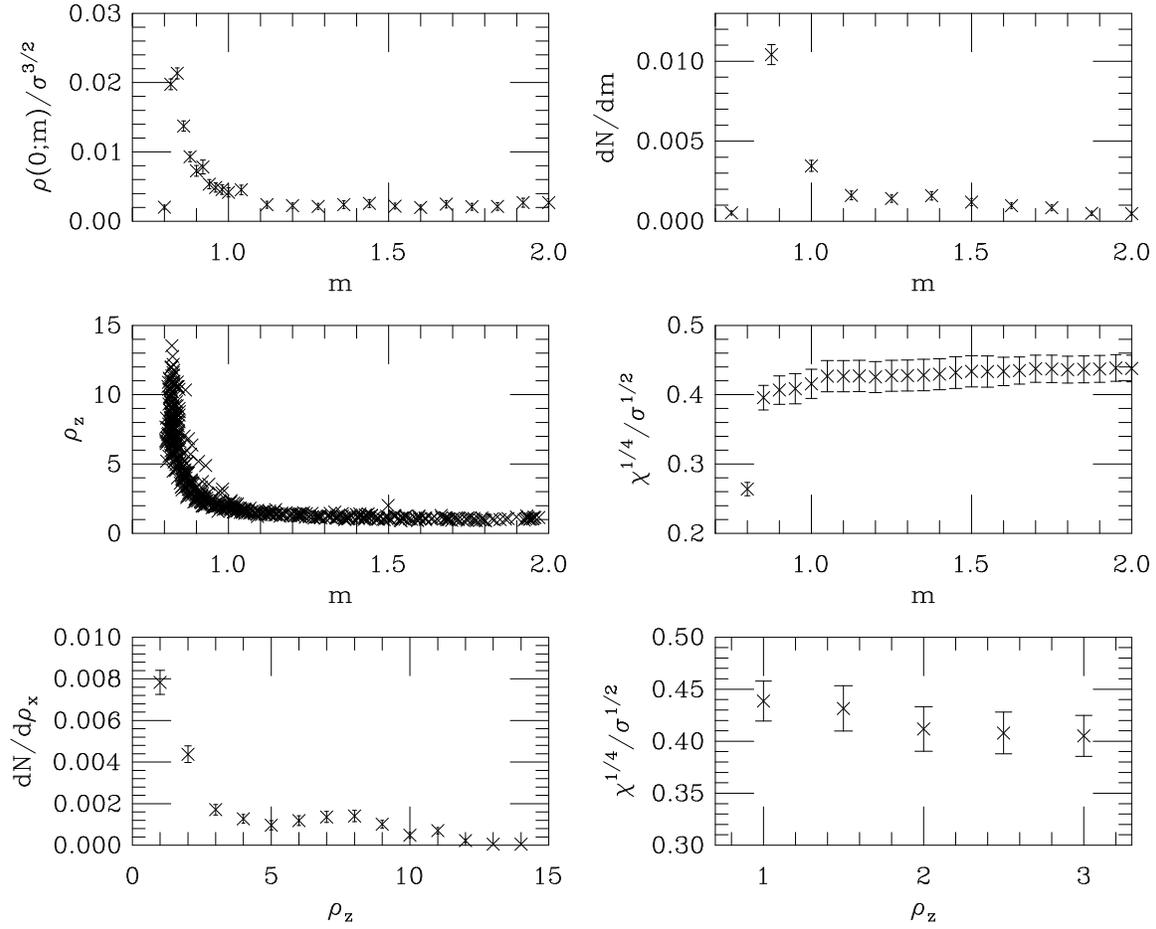}}
\end{center}
\vspace*{-10mm}
\caption{
  Detailed study for $\beta=6.0$, $16^3\times 32$.
}
\label{b6.0_all}
\end{figure*}

\section{SPECTRUM OF THE HERMITIAN WILSON--DIRAC OPERATOR}
\label{spec_flow}

The overlap formalism for constructing a chiral gauge theory on the
lattice~\cite{over} provides a natural definition of the index, $I$,
of the associated chiral Dirac operator. The index is equal to half
the difference of negative and positive eigenvalues of the hermitian
Wilson-Dirac operator
$\ham_L(m) = \gamma_5 W(-m)$
where $W(m)$ is the usual Wilson--Dirac operator 
(we will use negative sign for the mass term throughout).

On the lattice, because of the additive mass renormalization, the
crossings of zero occur at positive $m$ and spread out in $m$.  
It is easy to see, that no eigenvalues of $\ham_L(m)$ can be zero for $m<0$.
Since in the free case the first doublers become massless at $m=2$ we
restrict ourselves to $m < 2$.  A simple way to compute the index $I$
is to compute the lowest eigenvalues of $\ham_L(m)$ at some suitably
small $m$ before any crossings of zero occurred.
Then $m$ is slowly varied and the number and direction of zero
crossings are tracked.  The net number at some $m_t$ is the index of
the overlap chiral Dirac operator.

%We use the Ritz functional~\cite{ritz} to obtain the 10 lowest eigenvalues
%in the spectrum of $\ham^2_L(m)$, and thus the 10 eigenvalues of
%$\ham_L(m)$ closest to zero. In addition to giving accurate estimates for
%the low lying eigenvalues, the Ritz functional method also gives the
%corresponding eigenvectors. Using these eigenvectors one can use first
%order perturbation theory to interpolate the eigenvalues between successive
%mass points. This enables one to obtain the spectral flow of $\ham_L(m)$ as
%a continuous function in $m$ from the computation at suitably spaced
%values.

We have applied this procedure to compute the index, which we take as
the definition of topological charge on the lattice, of various gauge
field ensembles with gauge group SU(3) including pure gauge, Symanzik
improved pure gauge, and two dynamical flavor generated gauge
fields~\cite{su3_top}, and recently pure gauge SU(2).  
We found that
the zero crossings start occurring at some, ensemble dependent,
$m_1>0$ and continue occurring for all $m$ in $m_1<m<2$ in
sufficiently large lattice volumes. 
Hence, the spectral gap is closed in the entire range.

To further investigate the zero level crossings, we compute the
density of zero eigenvalues $\rho(0;m)$ for $\ham_L(m)$ by fitting
the integrated density $\int_0^\lambda \rho(\lambda') \,d\lambda'$
%
%\begin{equation}
%\int_0^\lambda \rho(\lambda') \,d\lambda' = 
%  \rho(0)\lambda + {\frac{1}{2}}\rho_1\lambda^2 + \ldots
%\end{equation}
%
to a line through the origin for some cutoff $\lambda$.  By studying
the scaling of $\rho(0;m)$ for quenched $SU(3)$, we conclude that the
density of zero eigenvalues falls exponentially in the inverse lattice
spacing and only vanishes in the continuum limit (see Figure~6 of
Narayanan's talk~\cite{rnaray_lat98}). We will observe latter that
these zeros are due to small localized modes.  More discussion can be
found in the talk of Narayanan~\cite{rnaray_lat98}.

\section{TOPOLOGY AND SMALL ZERO MODES}
\label{topology}

Using the index of the chiral Dirac operator as our 
(fermionic) definition of the topological charge of the gauge field
background, we obtain the topological susceptibility $\chi$ as a
function of $m_t$.  All of the gauge ensembles studied have the
general characteristic that the susceptibility rises sharply in the
region where $\rho(0;m)$ is peaked, and then it essentially flattens out
in the region where $\rho(0;m)$ is small~\cite{su3_top}.  
% THIS COULD GO TO SAVE SPACE. NOT VERY CLEAR.
%This is the case also in the presence of the SW term.
% Commented to save space !!
%This is
%the case also in the presence of the SW term, for the improved gauge
%action and for the ensembles that contain dynamical fermions.

\begin{figure}
%\vspace*{-10mm} \hspace*{-0cm}
\begin{center}
\epsfxsize = 0.45\textwidth
\centerline{\epsfbox[100 130 550 590]{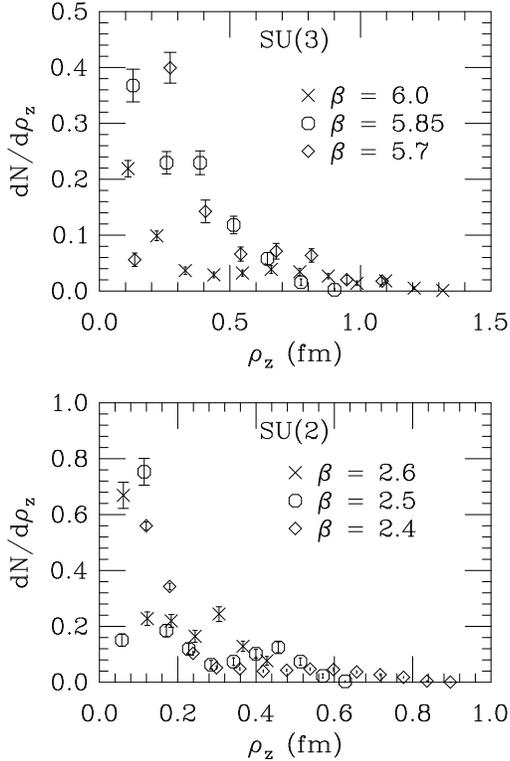}}
%\centerline{\ewxy{dN_all.ps}{100mm}}
%\leavevmode\epsffile{rho_talk.ps}
\end{center}
%\vspace*{-5mm}
\caption{
  Zero mode size distribution.
}
\label{dN_all}
\end{figure}

We define a size of the eigenvector associated with the level crossing
zero mode
\begin{displaymath}
\rho_z = {\frac{1}{2}}  {{\sum_t f(t)} \over {f_{\rm max}}}
\quad
%\end{equation}
%\begin{equation}
f(t) = 
\sum_{\vec k} {\rm tr}(\phi^\dag({\vec k},t)\phi({\vec k},t))
\nonumber
\end{displaymath}
motivated by the t'Hooft zero mode \ $\rho_z^2 / 2(t^2 + \rho_z^2)^{3/2}$
where $\phi_k$ are the eigenvectors of $\ham_L$. Another definition
based on the second moment of $f(t)$
was used in Ref.~\cite{su3_top}. We should emphasize
that we look only at the sizes of eigenmodes that cross, and only
close to the crossing point. Only then can we expect to get a good
estimate of the localization size inspired by the 't Hooft zero mode.

%$\beta$ & size & $N_{\rm conf}$ & $\chi^{1/4}/\sqrt\sigma$ \\
%$6.0$ & $16^3\times 32$ & 75 & 0.440(19) \\
%$5.85$ & $8^3\times 16$ & 200 & 0.450(11) \\
%$5.7$ & $8^3\times 16$ & 50 & 0.437(22) \\
%$2.6$ & $16^4$ & 400 & 0.519(11) \\
%$2.5$ & $16^4$ & 100 & 0.527(19) \\
%$2.4$ & $16^4$ & 200 & 0.501(12) \\

\begin{table}[t]
%\begin{center}
\small
\begin{tabular}{|c||l|c|r|l|}
\hline
&$\beta$ & size & $N_{\rm conf}$ & $\chi^{1/4}$(MeV) \\
\hline
&$6.0$ & $16^3\times 32$ & 75 & 194(10) \\
$SU(3)$ &$5.85$ & $8^3\times 16$ & 200 & 198(05) \\
&$5.7$ & $8^3\times 16$ & 50 & 193(10) \\
\hline
&$2.6$ & $16^4$ & 400 & 229(05) \\
$SU(2)$ &$2.5$ & $16^4$ & 100 & 232(10) \\
&$2.4$ & $16^4$ & 200 & 220(06) \\
\hline
\end{tabular}
%\end{center}
\caption{Topological susceptibility and parameters.}
%for $SU(3)$ and $SU(2)$.}
\label{tab:results}
\end{table}

%We found a monotonic relation between the crossing point
%$m$ and the size of the corresponding zero mode, with the crossings
%for larger objects occurring at smaller $m$. All crossings for objects
%with size larger than about two lattice spacings occurred within a
%small region of $m>m_1$. All later zero crossings correspond to small
%objects of size about one to two lattice spacings. These small objects
%do not seem to have physical effects and, for example, do not affect
%the topological susceptibility~\cite{su3_top}.

We show in Figure~\ref{b6.0_all} a detailed study of the $\beta=6.0$,
$16^3\times32$ pure gauge ensemble. On the first line is shown the
density of zero eigenvalues $\rho(0;m)$ and the number of crossings in
each mass bin. Since there are a nonzero number of crossings, we see
that $\rho(0;m)$ does indeed measure zero eigenvalues, and not just
small eigenvalues near zero. We also see that $\rho(0;m)$ rises
sharply in $m$, then falls to a nonzero value where there is a small
number of zero level crossings.

On the second line of Figure~\ref{b6.0_all}, we show the size of the
zero modes $\rho_z(m)$. The modes are large near $m_1$ where
$\rho(0;m)$ is large, then $\rho_z$ drops to about $1$ or $2$ lattice
spacings up to $m=2$. We see that the corresponding $\chi$
% topological susceptibility 
rises sharply when $\rho_z$ is large for $m$ near
$m_1$, then is quite stable when $\rho_z$ is small. This result show
that while the index, $I$, of the field is $m$ dependent, 
$\chi$
%the topological susceptibility 
(a physical quantity) is independent of
the contribution from the small modes for $m > 1$.

To further clarify the relative contribution of the zero modes, in the
last line of Figure~\ref{b6.0_all} the zero mode size distribution is
shown which peaks for $\rho_z \le 2$. In the adjacent graph, 
% the topological susceptibility 
$\chi$, here defined by the contribution of zero modes of size $\rho_z$ and
larger, is stable when $\rho_z \le 2$. Hence, the small modes do not
affect the estimate of $\chi$. Our estimates of $\chi$ are shown in
Table~\ref{tab:results} where we use the string tension value
$\sqrt\sigma = 440$(MeV) to set the scale. Our results are in rough
agreement with other groups~\cite{boulder,ETH,Pisa}.

In Figure~\ref{dN_all} we plot the zero mode size distribution for
several $\beta$'s for $SU(3)$ and $SU(2)$ which is of interest for the
instanton liquid model. We find the
distributions always peaks for sizes about $1$ - $2$ lattice spacings.
For $SU(3)$, we see small secondary peaks at $\sim 0.65$fm for
$\beta=6.0$ and $5.7$. However, there is only a shoulder for $5.85$ at
$0.3$fm. For $SU(2)$ which are all the same lattice size, we find the
secondary peaks shift to smaller $\rho_z$ for increasing
$\beta$. These secondary peaks do not occur at any general physical
radius and are consistent with a finite volume effect.

%This research was supported by DOE contracts 
%DE-FG05-85ER250000 and DE-FG05-96ER40979.
%Computations were performed on the QCDSP, CM-2 and the 
%workstation cluster at SCRI.

%
%
%%%%%%%%%%%%   references  %%%%%%%%%%%%%%%%%%%%%%%%
%

\end{document}